\documentclass{article}
\usepackage{amsmath}
\usepackage{harvard}

\input{tcilatex}

\begin{document}

\title{5D-Black Hole Solution in Einstein-Yang-Mills-Gauss-Bonnet Theory}
\author{S. Habib Mazharimousavi$^{\ast }$ and M. Halilsoy$^{\dagger }$ \\
%EndAName
Department of Physics, Eastern Mediterranean University,\\
G. Magusa, north Cyprus, Mersin-10, Turkey\\
$^{\ast }$habib.mazhari@emu.edu.tr\\
$^{\dagger }$mustafa.halilsoy@emu.edu.tr}
\maketitle

\begin{abstract}
By adopting the 5D version of the Wu-Yang Ansatz we present in closed form a
black hole solution in the Einstein-Yang-Mills-Gauss-Bonnet (EYMGB) theory.
In the EYM limit, we recover the 5D black hole solution already known.
\end{abstract}

\bigskip The 5D line element is chosen as 
\begin{equation}
ds^{2}=-f(r)\;dt^{2}+\frac{dr^{2}}{f(r)}+r^{2}d\;\Omega _{3}^{2}
\end{equation}%
in which the $S^{3}$ line element will be expressed in the alternative form%
\begin{equation}
d\Omega _{3}^{2}=\frac{1}{4}\left( d\theta ^{2}+d\phi ^{2}+d\psi ^{2}-2\cos
\theta \;d\phi \;d\psi \right)
\end{equation}%
where 
\begin{equation*}
0\leq \theta \leq \pi ,0\leq \phi ,\psi \leq 2\pi .
\end{equation*}

This form has a nearly Eulidean appearance with the advantage of admitting
two explicit killing vectors $\partial _{\phi }$ and $\partial _{\psi }$. We
introduce the Wu-Yang Ansatz [1,2] in 5D as%
\begin{eqnarray}
A^{i} &=&\frac{Q}{r^{2}}\epsilon _{jk}^{i}x^{k}dx^{j} \\
A^{i+3} &=&\frac{Q}{r^{2}}\left( x^{i}dw-wdx^{i}\right)  \notag \\
&&\left( i,j,k=1,2,3\right)  \notag
\end{eqnarray}%
in which $r^{2}=x^{2}+y^{2}+z^{2}+w^{2}$ and $Q$ is the only non-zero gauge
charge$.$ The latter coordinates are expressed in terms of the Euler angles%
\begin{eqnarray}
x+iy &=&r\sin \left( \frac{\theta }{2}\right) .\exp i\left( \frac{\phi +\psi 
}{2}\right) \\
z+iw &=&r\cos \left( \frac{\theta }{2}\right) .\exp i\left( \frac{\phi -\psi 
}{2}\right) .  \notag
\end{eqnarray}

It is observed that the reduction $S^{3}\rightarrow S^{2}$ amounts to taking 
$\phi =\psi $ and $\theta \rightarrow 2\theta .$ Although the isomorphism $%
SO\left( 4\right) \equiv SO\left( 3\right) \times SO\left( 3\right) ,$
supports two independent sets of rotation matrices, this will not be our
strategy here. Instead, we shall parametrize both groups in terms of the
common Euler angles which implies mixing of the groups.

The gauge potential 1-forms in terms of the Euler angles have the following
explicit form%
\begin{eqnarray}
A^{1} &=&\frac{Q}{4}\left( \cos \psi \sin \theta \;d\phi +\cos \phi \sin
\theta \;d\psi +\left( \sin \phi +\sin \psi \right) \;d\theta \right) \\
A^{2} &=&\frac{Q}{4}\left( \sin \psi \sin \theta \;d\phi +\sin \phi \sin
\theta \;d\psi -\left( \cos \phi +\cos \psi \right) \;d\theta \right)  \notag
\\
A^{3} &=&-\frac{Q}{2}\sin ^{2}\left( \frac{\theta }{2}\right) \left( d\phi
+d\psi \right)  \notag \\
A^{4} &=&\frac{Q}{4}\left( \cos \psi \sin \theta \;d\phi -\cos \phi \sin
\theta \;d\psi -\left( \sin \phi -\sin \psi \right) \;d\theta \right)  \notag
\\
A^{5} &=&\frac{Q}{4}\left( \sin \psi \sin \theta \;d\phi -\sin \phi \sin
\theta \;d\psi +\left( \cos \phi -\cos \psi \right) \;d\theta \right)  \notag
\\
A^{6} &=&\frac{Q}{2}\cos ^{2}\left( \frac{\theta }{2}\right) \left( d\phi
-d\psi \right) .  \notag
\end{eqnarray}

The YM field 2-forms are defined as follow%
\begin{eqnarray}
F^{1} &=&dA^{1}+\frac{1}{Q}\left( A^{2}\wedge A^{3}+A^{5}\wedge A^{6}\right)
\\
F^{2} &=&dA^{2}+\frac{1}{Q}\left( A^{3}\wedge A^{1}+A^{6}\wedge A^{4}\right)
\notag \\
F^{3} &=&dA^{3}+\frac{1}{Q}\left( A^{1}\wedge A^{2}+A^{4}\wedge A^{5}\right)
\notag \\
F^{4} &=&dA^{4}+\frac{1}{Q}\left( A^{2}\wedge A^{6}+A^{5}\wedge A^{3}\right)
\notag \\
F^{5} &=&dA^{5}+\frac{1}{Q}\left( A^{6}\wedge A^{1}+A^{3}\wedge A^{4}\right)
\notag \\
F^{6} &=&dA^{6}+\frac{1}{Q}\left( A^{4}\wedge A^{2}+A^{1}\wedge A^{5}\right)
.  \notag
\end{eqnarray}

We note that our notation follows the standard exterior differential forms,
namely $d$ stands for the exterior derivative while $\wedge $ stands for the
wedge product. The hodge star $\ast $ in the sequel will be used to
represent duality [3].

The integrability conditions 
\begin{equation}
dF^{1}+\frac{1}{Q}\left( A^{2}\wedge F^{3}-A^{3}\wedge F^{2}+A^{5}\wedge
F^{6}-A^{6}\wedge F^{5}\right) =0
\end{equation}%
plus other five similar equations, are easily satisfied by using (6). The YM
equations 
\begin{eqnarray}
d^{\ast }F^{1}+\frac{1}{Q}\left( A^{2}\wedge ^{\ast }F^{3}-A^{3}\wedge
^{\ast }F^{2}+A^{5}\wedge ^{\ast }F^{6}-A^{6}\wedge ^{\ast }F^{5}\right) &=&0
\\
d^{\ast }F^{2}+\frac{1}{Q}\left( A^{3}\wedge ^{\ast }F^{1}-A^{1}\wedge
^{\ast }F^{3}+A^{6}\wedge ^{\ast }F^{4}-A^{4}\wedge ^{\ast }F^{6}\right) &=&0
\notag \\
d^{\ast }F^{3}+\frac{1}{Q}\left( A^{1}\wedge ^{\ast }F^{2}-A^{2}\wedge
^{\ast }F^{1}+A^{4}\wedge ^{\ast }F^{5}-A^{5}\wedge ^{\ast }F^{4}\right) &=&0
\notag \\
d^{\ast }F^{4}+\frac{1}{Q}\left( A^{2}\wedge ^{\ast }F^{6}-A^{6}\wedge
^{\ast }F^{2}+A^{5}\wedge ^{\ast }F^{3}-A^{3}\wedge ^{\ast }F^{5}\right) &=&0
\notag \\
d^{\ast }F^{5}+\frac{1}{Q}\left( A^{6}\wedge ^{\ast }F^{1}-A^{1}\wedge
^{\ast }F^{6}+A^{3}\wedge ^{\ast }F^{4}-A^{4}\wedge ^{\ast }F^{3}\right) &=&0
\notag \\
d^{\ast }F^{6}+\frac{1}{Q}\left( A^{4}\wedge ^{\ast }F^{2}-A^{2}\wedge
^{\ast }F^{4}+A^{1}\wedge ^{\ast }F^{5}-A^{5}\wedge ^{\ast }F^{1}\right) &=&0
\notag
\end{eqnarray}%
are all satisfied.

The energy-momentum tensor 
\begin{equation}
T_{\mu \nu }=2F_{\;\mu }^{i\;\;\alpha }F_{\;\nu \alpha }^{i}-\frac{1}{2}%
g_{\mu \nu }F_{\;\alpha \beta }^{i}F^{i\;\alpha \beta }
\end{equation}%
where $F_{\;\alpha \beta }^{i}F^{i\;\alpha \beta }=6Q^{2}/r^{4},$ has the
non-zero components 
\begin{eqnarray}
T_{tt} &=&\frac{3Q^{2}f(r)}{r^{4}} \\
T_{rr} &=&-\frac{3Q^{2}}{r^{4}f(r)}  \notag \\
T_{\theta \theta } &=&T_{\phi \phi }=T_{\psi \psi }=\frac{Q^{2}}{4r^{2}} 
\notag \\
T_{\phi \psi } &=&T_{\psi \phi }=-\frac{Q^{2}}{4r^{2}}\cos \theta .  \notag
\end{eqnarray}

The EYMGB equations[4] 
\begin{gather}
G_{\mu \nu }-\alpha \left\{ \frac{1}{2}g_{\mu \nu }\left( R_{\kappa \lambda
\rho \sigma }R^{\kappa \lambda \rho \sigma }-4R_{\rho \sigma }R^{\rho \sigma
}+R^{2}\right) -2RR_{\mu \nu }\right.  \notag \\
+\left. 4R_{\mu \lambda }R_{\nu }^{\;\lambda }+4R^{\rho \sigma }R_{\mu \rho
\nu \sigma }-R_{\mu }^{\;\rho \sigma \lambda }R_{\nu \rho \sigma \lambda
}\right\} =T_{\mu \nu },
\end{gather}%
reduce to the simple set of equations 
\begin{gather}
2r\left( -\frac{r^{2}}{4}+\alpha \left( f\left( r\right) -1\right) \right)
f^{\prime }\left( r\right) +r^{2}\left( 1-f\left( r\right) \right) -Q^{2}=0
\\
r^{2}\left( \frac{r^{2}}{2}+2\alpha \left( 1-f\left( r\right) \right)
\right) f^{\prime \prime }\left( r\right) +2r^{3}f^{\prime }\left( r\right)
-2\alpha r^{2}f^{\prime }\left( r\right) ^{2}+\left( f\left( r\right)
-1\right) r^{2}-Q^{2}=0  \notag
\end{gather}%
in which a prime denotes derivative with respect to $r$.

This set admits the solution 
\begin{equation}
f\left( r\right) =1+\frac{r^{2}}{4\alpha }\pm \sqrt{\left( \frac{r^{2}}{%
4\alpha }\right) ^{2}+(1+\frac{m}{2\alpha })+\frac{Q^{2}\ln \left( r\right) 
}{\alpha }}
\end{equation}%
\bigskip in which $m$ is the usual integration constant to be identified as
mass. We notice that under the limit $\alpha \rightarrow 0,$ and the $(-)$
sign (i.e. EYM limit) the solution reduces to [5] 
\begin{equation}
f\left( r\right) =1-\frac{m}{r^{2}}-\frac{2Q^{2}}{r^{2}}\ln \left( r\right) .
\end{equation}

The difference of this solution from the 5D Reissner-Nordstrom solution
requires no comment.

By using (10) the energy density is 
\begin{equation}
\epsilon =-g^{tt}T_{tt}=\frac{3Q^{2}}{r^{4}}
\end{equation}%
whose integral diverges logarithmically as in the Reissner-Nordstrom case.
The surface gravity, $\kappa $ defined by [6] (note that for our purpose we
are choosing the (-) sign and $\alpha \geq 0$ in (13)) 
\begin{equation}
\kappa ^{2}=-\frac{1}{4}g^{tt}g^{ij}g_{tt,i}\;g_{tt,j}
\end{equation}%
has the form 
\begin{equation}
\kappa =\left\vert \frac{1}{2}f^{\prime }\left( r_{+}\right) \right\vert
=\left\vert \frac{r}{4\alpha }-\left( \frac{r_{+}^{3}}{4\alpha ^{2}}+\frac{%
Q^{2}}{r_{+}\alpha }\right) \frac{1}{\sqrt{\Delta }}\right\vert \text{ \ }
\end{equation}%
where%
\begin{equation}
\Delta =\frac{r_{+}^{4}}{\alpha ^{2}}+16+\frac{8m}{\alpha }+\frac{16Q^{2}\ln
\left( r_{+}\right) }{\alpha }
\end{equation}%
and $r_{+}$ is the radius of the event horizon which is the grater root of $%
f\left( r\right) =0.$

This can be reduced to the following simple equation%
\begin{equation}
r^{2}-m-2Q^{2}\ln \left( r\right) =0,
\end{equation}%
which is $\alpha $ independent. It is remarkable to observe that $\alpha $
does not change the radius of the event horizon.

To go further, let us take $m=1,$ with $Q<1,$ then one can easily show that
the particular radius of the event horizon $r_{+}$ is equal to $1$ and
consequently 
\begin{equation}
\kappa =\frac{1-Q^{2}}{4\alpha +1}.
\end{equation}

Clearly at the EYM limit$\left( \text{i.e.}.\text{ }\alpha \rightarrow
0\right) $ $\kappa =1-Q^{2}$ and asymptotically when $\alpha \rightarrow
\infty $ , $\kappa \rightarrow 0$ which states that the space is flat. The
associated Hawking temperature is given by 
\begin{equation}
T_{H}=\frac{\kappa }{2\pi }=\frac{1}{2\pi }\frac{1-Q^{2}}{4\alpha +1}
\end{equation}%
in natural unit $c=G=\hbar =k=1.$\ \ \ 

\bigskip The expression (13) suggests that the square root term must be
positive, this restricts our choice of the Gauss-Bonnet parameter $\alpha $
to certain limits. In order to get rid of the negative sign in the square
root for any arbitrary $\alpha ,$ one can shift the origin to the largest
root of the square root term [4].

As a final remark we wish to express optimism that in a similar manner it is
possible to construct analogous solutions for the EYMGB equations in higher
dimensions. This all amounts to defining appropriate gauge potentials and
overcoming the tedious calculations.

\end{document}